\documentclass[aps,english,superscriptaddress,twocolumn,pra]{revtex4-1}
\usepackage{amsfonts}
\usepackage{amssymb}
\usepackage{amsmath}
\usepackage{graphicx}
\usepackage{epsfig}
\usepackage{makecell}
\usepackage{enumerate}
\usepackage{mathrsfs}
\usepackage{color,xcolor}

\begin{document}

\title{Two-dimensional anisotropic non-Hermitian Lieb lattice}
\author{L. C. Xie}
\affiliation{School of Physics, Nankai University, Tianjin 300071, China}
\author{H. C. Wu}
\affiliation{School of Physics, Nankai University, Tianjin 300071, China}
\author{X. Z. Zhang}
\email{zhangxz@tjnu.edu.cn}
\affiliation{College of Physics and Materials Science, Tianjin Normal University, Tianjin
300387, China}
\author{L. Jin}
\email{jinliang@nankai.edu.cn}
\affiliation{School of Physics, Nankai University, Tianjin 300071, China}
\author{Z. Song}
\affiliation{School of Physics, Nankai University, Tianjin 300071, China}

\begin{abstract}
We study an anisotropic two-dimensional non-Hermitian Lieb lattice, where
the staggered gain and loss present in the horizontal and vertical
directions, respectively. The intra-cell nonreciprocal coupling generates
magnetic flux enclosed in the unit cell of the Lieb lattice and creates
nontrivial topology. The active and dissipative topological edge states are
along the horizontal and vertical directions, respectively. The
two-dimensional non-Hermitian Lieb lattice also supports passive topological
corner state. At appropriate magnetic flux, the non-Hermiticity can alter
the corner state from one corner to the opposite corner as the
non-Hermiticity increases. The gapless phase of the Lieb lattice is
characterized by different configurations of exceptional points in the
Brillouin zone. The topology of the anisotropic non-Hermitian Lieb lattices
can be verified in many experimental platforms including the optical
waveguide lattices, photonic crystals, and electronic circuits.
\end{abstract}

\maketitle

\section{Introduction}

Over the last two decades, the ubiquitous effect of dissipation has proven
to induce astonishing non-Hermitian features, rather than just being an
inescapable nuisance. Within this young field, the treasure hunt is
sprouting into fascinating new directions ranging from the complex optical
media \cite{Gupta2021}, non-equilibrium open systems with gain and/or loss
\cite%
{Hu2014,Kaiser2014,Mitrano2016,Cantaluppi2018,Sato2019,McIver2020,Booker2020,Tindall2020}
to strongly correlated systems as a result of finite-lifetime quasiparticles
\cite%
{Zhang2017a,Shen2018,Yoshida2018,Zhang2020a,Zhang2020b,Nakagawa2020,Nakagawa2021}%
. Most recently, topological characterization and dynamic control of
non-Hermitian models are hot areas of research \cite%
{Longhi2019,Kawabata2019b,Zhang2019,Luo2019,Yuce2019,Du2019,Ge2019,Lieu2019,Jiang2020,Xu2020,Wu2020a,Liu2020,Xu2020a,Wojcik2020,Kawabata2020,Cui2020,Li2020,Zeng2020,Wu2020,Guo2020,Zhang2020c,Zhang2020d,Zhang2020e,BgWang2020,Yokomizo2020,Wang2021a,Lang2021,Xu2021,Rivero2021,Hou2021,Longhi2021,Schiffer2021,Zhang2021}%
. Among the most relevant features observed in non-Hermitian systems, the
appearance of an interface significantly alters the entire spectrum, leading
to the exponential localization of all eigenmodes at the interface, which
goes beyond the expectations for Hermitian systems. This unique
non-Hermitian effect is dubbed as non-Hermitian skin effect \cite{Yao2018a}.
As a consequence, the conventional bulk-boundary correspondence breaks down.
A correct description requires one to extend Bloch band theory into the
generalized Brillouin zone (BZ) \cite{Yao2018a,Zhang2020f,Yang2020}.
Inspired by these exciting advances, different generalized versions of
bulk-boundary correspondence based on redefining the bulk topological
indices to incorporate the impact of non-Hermitian skin effect have been
proposed \cite%
{Imura2019,Song2019a,Song2019b,Yokomizo2019,Jiang2019,Yi2020,Longhi2019a,Wang2019,Borgnia2020,Longhi2020,Okuma2020,Zhu2020b,Zeng2020a,Kawabata2020a,Silberstein2020,Wang2020,Koch2020}%
. Furthermore, non-Hermitian skin effect itself is also a topological effect
manifested by the spectral winding on the complex energy plane with a
reference energy. Importantly, the non-Hermitian topological phenomena have
been observed experimentally in various experimental platforms including the
optical waveguide lattices, photonic crystals, and electronic circuits \cite%
{Pocock2019,Ezawa2019,Zhu2020a,Li2020a,Yin2020,Qin2020,Sone2020,Xiao2020,Ota2020,Sun2020,Zhu2020,Zhang2020g,Wang2021,Parto2021}%
.

The exceptional point (EP) uniquely presents in the non-Hermitian
Hamiltonian \cite{Berry2004,Heiss2012,Lee2016,Miri2019}. At the EPs, the
eigenstate coalescence occurs. The order of the EP depends on the geometric
multiplicity of the corresponding eigenvalue. In general, the band touching
induces EPs in one-dimensional system. Intriguingly, the EPs are
topologically stable at generic points in the BZ in the sense that they will
not disappear suddenly, but move, split, and merge in the BZ until merging
in pairs.

In this paper, we study an anisotropic two-dimensional (2D) non-Hermitian
Lieb lattice. The gain and loss present in the horizontal and vertical
directions, respectively; and induce the active and dissipative topological
edge states. The Lieb lattice supports passive topological corner state,
which is created at the appropriate cooperation between the non-Hermiticity
and the magnetic flux. The non-Hermitian term affects the corner states in a
subtle way: when the intra-cell nonreciprocal coupling induced magnetic flux
is $\pi/2$, the corner states are always localized at one corner; however,
when the intra-cell nonreciprocal coupling induced magnetic flux is $-\pi/2$%
, the corner states slowly evolve into the opposite corner as the
non-Hermiticity increases. The type and configuration of EPs in the BZ
distinguish the gapless phases. Our findings shed light on the influence of
non-Hermiticity for the 2D Lieb lattice.

The remainder of the paper is organized as follows. In Sec.~\ref{II}, we
introduce the anisotropic 2D non-Hermitian Lieb lattice. Section~\ref{III}
presents the topology of non-Hermitian Lieb lattice and the topological edge
states in the gapped phase. The corner states under the influence of
non-Hermiticity are elaborated. Section~\ref{IV} highlights the enriched
gapless band structures characterized by the EPs. Our conclusion is
summarized in Sec.~\ref{V}.

\section{2D non-Hermitian Lieb lattice}

\label{II}

\begin{figure}[t]
\setlength{\abovecaptionskip}{0cm} \negthinspace
\includegraphics[bb=25 305
540 540, width=15.8cm, angle=0,clip]{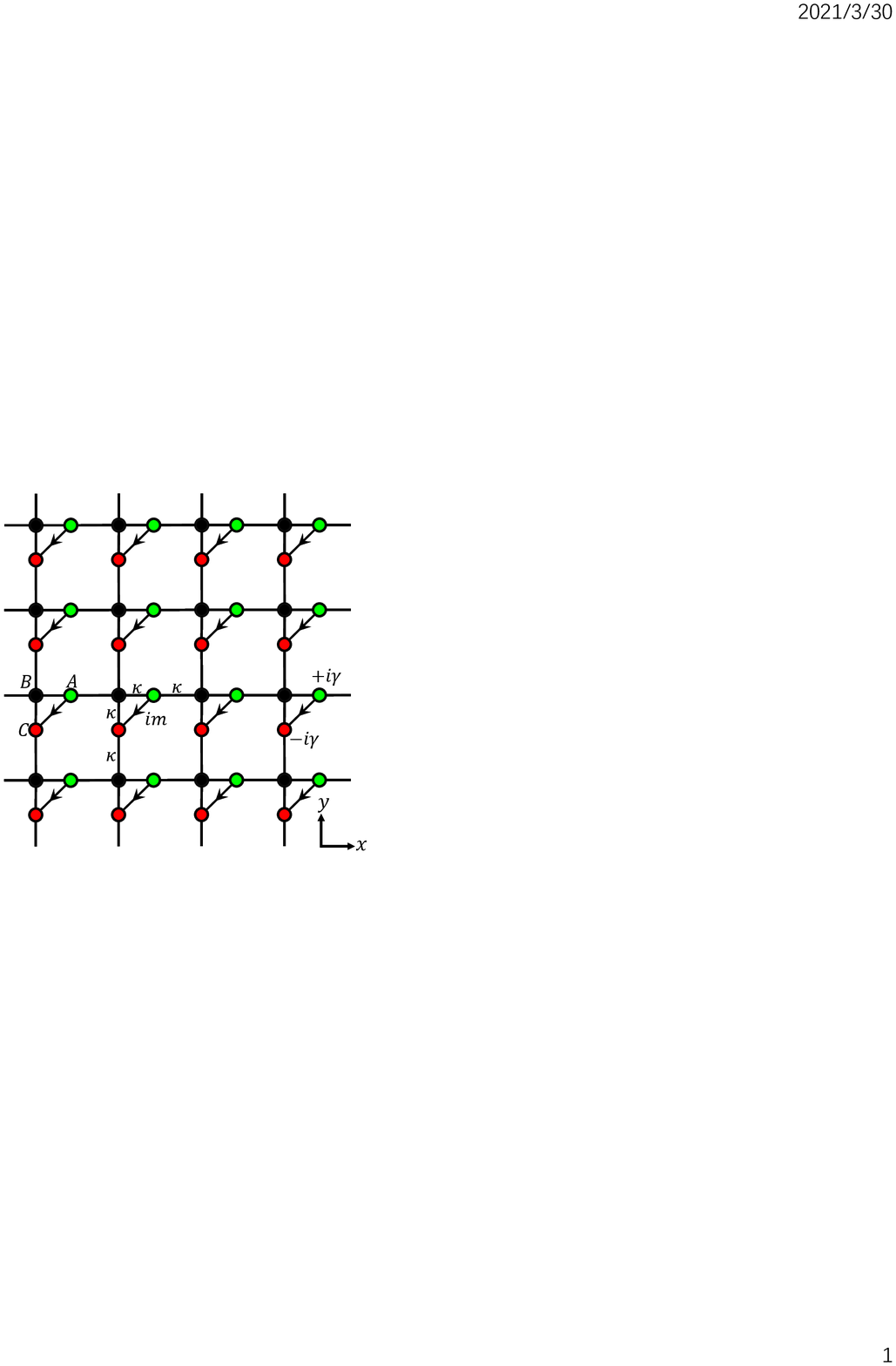}
\caption{Schematic illustration of the 2D non-Hermitian Lieb lattice. The
system consists of three sublattices denoted by green, black and red solid
circles, respectively. The non-Hermiticity arises from the gain in red and loss in green. The black
arrow indicates nonreciprocal coupling and each unit cell is threaded by a magnetic flux.}
\label{Lattice}
\end{figure}

We consider an anisotropic 2D non-Hermitian Lieb lattice. The Hamiltonian is
written in the form of
\begin{align}
H& =\sum\limits_{l,n}(\kappa A_{l,n}^{\dagger }B_{l,n}+\kappa
B_{l,n}^{\dagger }C_{l,n}+imC_{l,n}^{\dagger }A_{l,n}  \notag \\
& +\kappa A_{l,n}^{\dagger }B_{l+1,n}+\kappa A_{l,n}^{\dagger }B_{l,n+1})+%
\mathrm{H.c.}  \notag \\
& +i\gamma (A_{l,n}^{\dagger }A_{l,n}-C_{l,n}^{\dagger }C_{l,n}),  \label{H}
\end{align}%
where $A^{\dagger }$, $B^{\dagger }$, and $C^{\dagger }$ ($A$, $B$, and $C$)
denote the creation (annihilation) operators for the three sublattices in
each unit cell. The non-Hermitian Lieb lattice is schematically illustrated
in Fig. \ref{Lattice}. $\kappa $ and $m$ are the coupling strengths. The
intra-cell nonreciprocal coupling $im$ between the sublattices $A$ and $C$
induces effective magnetic flux $\pi /2$ enclosed in the unit cell. The
non-Hermiticity $\gamma $ originates from the balanced gain and loss that
presented in the sublattices $A$ and $C$ in the horizontal and vertical
directions, respectively. The proposed non-Hermitian Lieb lattice can be
experimentally implemented in the platforms of ultracold atomic gas in
optical lattices, photonic crystals, and coupled resonators based on the
nowadays technology \cite%
{YXXiao2020,Goldman2011,Guzman-Silva2014,Vicencio2015,Taie2015,Mukherjee2015,Whittaker2018}%
.

\begin{figure}[t]
\includegraphics[bb=-10 0 480 440,width=7cm,clip]{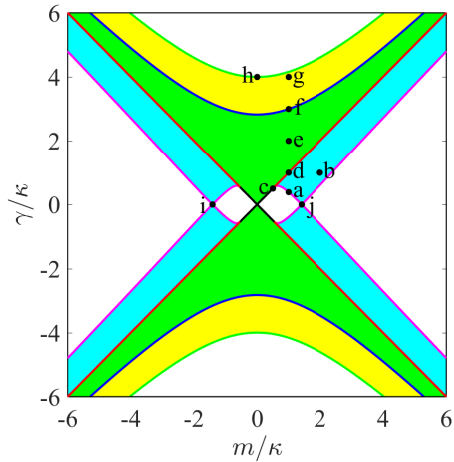} \setlength{		%
\abovecaptionskip}{0.5cm} \negthinspace
\caption{Phase diagram in the $m$-$\protect\gamma $ parameter space.
The color regions show the three gapless phases distinguished by the EPs in the BZ.
In white regions show the gapped phases distinguished by the Chern numbers.} %
\label{PhaseD1}
\end{figure}

Applying the Fourier transformation $a_{l,n}=N^{-1}\sum_{\mathbf{k}}e^{i%
\mathbf{k\cdot r}}a_{\mathbf{k}}$ $(a=A,B,C)$ for the three sublattices, the
Hamiltonian $H$ rewritten in the Nambu representation reads%
\begin{equation}
H=\sum_{\mathbf{k}}H\left( \mathbf{k}\right) =\sum_{\mathbf{k}}\psi _{%
\mathbf{k}}^{\dagger }h\left( \mathbf{k}\right) \psi _{\mathbf{k}},
\end{equation}%
where the basis is $\psi _{\mathbf{k}}=[A_{\mathbf{k}},B_{\mathbf{k}},C_{%
\mathbf{k}}]^{T}$ and $h\left( \mathbf{k}\right) $ is a $3\times 3$ matrix
\begin{equation}
h\left( \mathbf{k}\right) =\left(
\begin{array}{ccc}
i\gamma & \kappa (e^{ik_{x}}+1) & -im \\
\kappa (e^{-ik_{x}}+1) & 0 & \kappa (e^{ik_{y}}+1) \\
im & \kappa (e^{-ik_{y}}+1) & -i\gamma%
\end{array}%
\right) .
\end{equation}%
The Hamiltonians in the momentum subspaces commute with each other $[H\left(
\mathbf{k}\right) ,H(\mathbf{k}^{\prime })]=0$.

At $m=\gamma =0$, the lattice is a standard Lieb lattice and supports a flat
band. In the absence of the gain and loss $\gamma =0$, the lattice has a
flat band and the band energy is tuned by the coupling strength $m$ \cite%
{Vidal1998,Wu2007,Chalker2010,Apaja2010,Bermudez2011,Bodyfelt2014,Julku2016,Shukla2018}%
. The wave transport in the flat band is completely suppressed because of
the momentum-independent dispersion relation, leading to a strong
localization of the eigenstates. This provides an ideal platform to
investigate various interesting strongly correlated phenomena \cite%
{Goda2006,Maksymenko2012,Peotta2015,Kauppila2016}.

The presence of gain and loss drastically alters the spectrum and results in
exotic phenomena comparing to the standard Lieb lattice. We investigate $%
h\left( \mathbf{k}\right) $ to show the insights of the non-Hermitian Lieb
lattice. The energy band properties of $h\left( \mathbf{k}\right) $ are
determined from solving the secular equation $\mathrm{det}[h\left( \mathbf{k}%
\right) -E\left( \mathbf{k}\right) I]=0$, where $I$ is the identity matrix.
The algebra after the basis transformation shows a cubic equation
\begin{equation}
E^{3}\left( \mathbf{k}\right) +p(h_{x},h_{y})E\left( \mathbf{k}\right)
+q(h_{x},h_{y})=0,  \label{cubic}
\end{equation}%
where%
\begin{eqnarray}
p(h_{x},h_{y}) &=&\gamma ^{2}-h_{x}^{2}-h_{y}^{2}-m^{2}, \\
q(h_{x},h_{y}) &=&i\gamma (h_{y}^{2}-h_{x}^{2})+2h_{x}h_{y}m\sin
[(k_{x}+k_{y})/2]
\end{eqnarray}%
with $h_{\epsilon }^{2}=|\kappa (e^{ik_{\epsilon }}+1)|^{2}$ ($\epsilon =x,$
$y$). The energy bands and eigenstates are straightforwardly obtained
through solving the cubic equation.

\section{Topological Phases}

\label{III}

\begin{figure}[tb]
\includegraphics[bb=-10 0 380 350,width=6.6cm,clip]{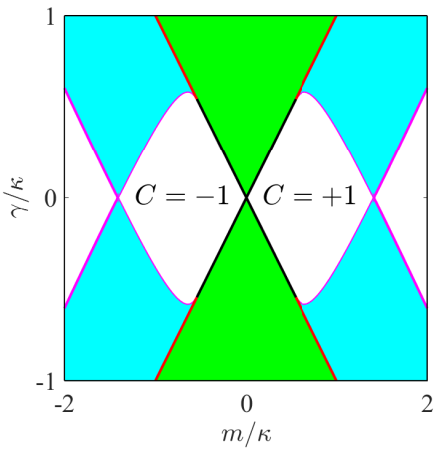} %
\setlength{\abovecaptionskip}{0.3cm}
\caption{Phase diagram in the $m$-$\protect\gamma$ parameter space focused
on the gapped regions. Here $\protect\kappa $ is taken as the unit without
loss of generality. In gapped regions, the system can be either in
topologically trivial or non-trivial phase characterized by the Chern number. $C=\pm 1$ indicates the Chen number of the lowest band of the topologically
non-trivial phase. Except for the white region marked with $C=1$, all other
white regions satisfy $C=0$. For the topologically non-trivial phase, edge
states exist for the system under OBC in the $x$ or $y$ direction.} \label%
{PhaseD2}
\end{figure}

\begin{figure}[tb]
\includegraphics[bb=0 20 700 620,width=8.8cm,clip]{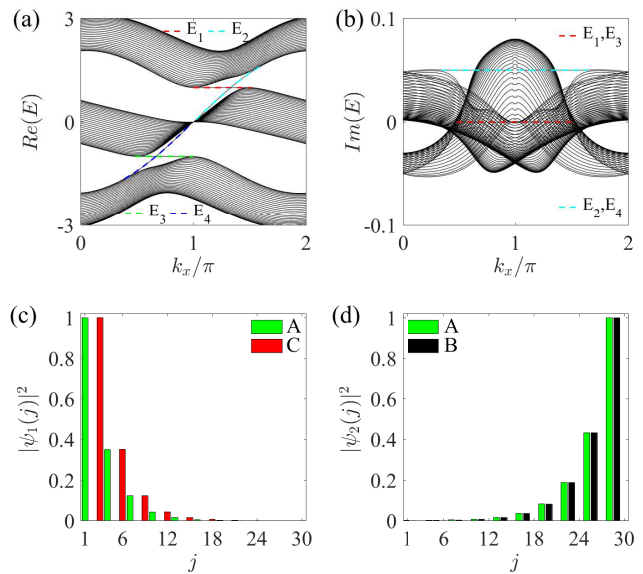}
\caption{Energy spectrum of the topologically nontrivial phase under PBC in
the $x$ direction, but under OBC in the $y$ direction. Here $\protect\kappa$
is taken as the unit without loss of generality, and the total number of the
unit cells $N$ is $10$. (a)-(b) $m=-1,$ $\protect\gamma =0.1$ depicts the
topologically nontrivial phase with gapless edge states. (c)-(d) the
probability distribution of $|\protect\psi _{1}\rangle $ and $|\protect\psi _{2}\rangle $ by supposing $k_{x}=1.4\protect\pi $. Different colour bars
indicate different sublattices, respectively.} \label{TNT1}
\end{figure}

In this section, we discuss the topological phase of the system based on the
non-Hermitian topological band theory \cite{Shen2018b}. For the
non-Hermitian Hamiltonian, the separable energy band means that any two
bands of the system are not degenerate at any point $\mathbf{k}$ in the
momentum space (i.e., $E_{\alpha }(\mathbf{k})\neq E_{\beta }(\mathbf{k})$, $%
\alpha $, $\beta $ are the band index). The system experiences topological
phase transition when the closing of separable bands occurs.

For separable bands, the topology of the anisotropic 2D non-Hermitian Lieb
lattice can be characterized by the Chern number. Clearly, there is a fixed
magnetic flux in the unit cell because of the nonreciprocal coupling
strength between the next nearest neighbor, which breaks the time-reversal
symmetry and ensures the existence of the Chern number. For nonzero Chern
number, the topological edge states exist for the system under OBC in the $x$
or $y$ direction; for zero Chern number, the system has trivial edge states
or no edge states when imposing OBC in the $x$ or $y$ direction.

\textit{Separable bands and Chern number. }Figure \ref{PhaseD2} is the phase
diagram. The nonzero Chern number are marked. The white regions except for
the topologically nontrivial region marked with $C=\pm 1$ are all
topologically trivial phases with the zero Chern number. The non-Hermiticity greatly affects the topological property of the Lieb lattice. When the non-Hermitian term is introduced, the gapped regions shrink and the gapless regions appear; a typical feature is the existence of the EPs. Moreover, the large non-Hermiticity destroys the nontrivial topology of the Lieb lattice as shown in the phase diagram Fig.~\ref{PhaseD1}.

The Chern number of the separable energy band is well-defined as the
integral of Berry curvature on the entire Brillouin zone.
\begin{equation}
C=\frac{1}{2\pi }\iint_{\mathrm{BZ}}\mathrm{d}k_{x}\mathrm{d}k_{y}\Omega ,
\end{equation}%
where $\Omega =\nabla \times \mathcal{A}$ with $\mathcal{A}=-i\left\langle
\varphi (\mathbf{k})\right\vert \nabla \left\vert \varphi (\mathbf{k}%
)\right\rangle $ and $\left\vert \varphi (\mathbf{k})\right\rangle $ is the
eigenstate of the band. The Chern number evaluated under PBC predicts the
number of gapless edge states of the system under OBC \cite{Shen2018b}. In
the topologically nontrivial phase, the Chern number for the lower band is $%
C=\pm 1$, the Chern number for the middle band is $C=0$, and the Chern
number for the upper band is $C=\mp 1$. In the topologically trivial phase,
the Chern number for the energy bands are all zeros.

\begin{figure}[tb]
\includegraphics[bb=0 20 700 620,width=8.8cm,clip]{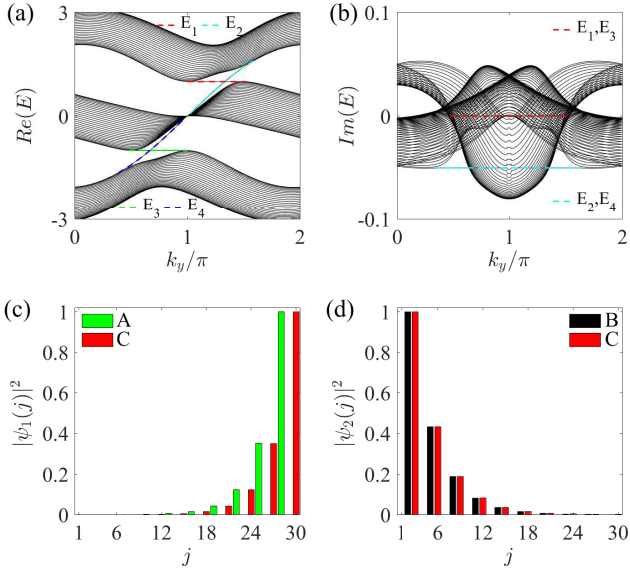}
\caption{Energy spectrum of the gapless edge states under PBC in
the $y$ direction, but under OBC in the $x$ direction. The (a) real and (b) imaginary parts of energy band at $m=-1,$ $\protect\gamma =0.1$.
The edge states (c) $|\protect\psi _{1}\rangle $ and (d) $|\protect\psi _{2}\rangle $
at $k_{y}=1.4\protect\pi $. Color sticks indicate the probabilities for the corresponding sublattices. The total number of unit cells $N$ is $10$ and $\protect\kappa=1$.}
\label{TNT2}
\end{figure}

As a consequence, the gapless edge states between the middle band and the
other bands exist for the Lieb lattice under OBC in the $x$ or the $y$
direction. Figure \ref{TNT1}(a) depicts the energy bands when the system
under PBC in the $x$ direction, but under OBC in the $y$ direction. Figure %
\ref{TNT2}(a) shows the energy band when PBC is applied in the $y$ direction
but OBC is applied in the $x$ direction. The non-Hermitian Hamiltonian $%
h\left( \mathbf{k}\right) $ in the momentum space has the particle-hole
symmetry, $\mathcal{T}_{-}h^{\ast }\left( \mathbf{k}\right) \mathcal{T}%
_{-}^{-1}=-h\left( -\mathbf{k}\right) $; the unitary operator $\mathcal{T}%
_{-}$ is defined as $\mathcal{T}_{-}=\mathrm{diag}\left( 1,-1,1\right) $.
Thus, the band energies satisfy $E_{\mathbf{k}}=-E_{-\mathbf{k}}^{\ast }$ as
reflected from Fig. \ref{TNT1}(a).

\textit{Edge states. }In the topologically nontrivial phase of the
non-Hermitian Lieb lattice, we examine the properties of the gapless edge
states. The edge states localized on the boundary in the $x$ and$\ y$
direction exhibit the gain and loss, respectively. We discuss the $C=-1$
region as an example. The energy spectra for the lattice under PBC in the $x$
direction and OBC in the $y$ direction are shown in Figs.~\ref{TNT1}(a)-\ref%
{TNT1}(b). Four edge states appear in pairs within the band gap.
Straightforward algebra shows that
\begin{eqnarray}
E_{1} &=&\sqrt{m^{2}-\gamma ^{2}}, \\
E_{2} &=&\sqrt{4\kappa ^{2}\cos ^{2}(k_{x}/2)-\gamma ^{2}/4}+i\gamma /2, \\
E_{3} &=&-\sqrt{m^{2}-\gamma ^{2}}, \\
E_{4} &=&-\sqrt{4\kappa ^{2}\cos ^{2}(k_{x}/2)-\gamma ^{2}/4}+i\gamma /2.
\end{eqnarray}%
The edge state energies between the middle and lowest bands satisfy $%
E_{3}=-E_{1}$ and $E_{4}=-E_{2}^{\ast }$ under the particle-hole symmetry. $%
E_{1}$ and $E_{3}$ are opposite in pair and independent of the momentum $%
k_{x}$. $E_{2}$ and $E_{4}$ are complex with constant gain rate $i\gamma /2$%
. The wavefunction of the edge state for $E_{\mu }$ is denoted as $|\psi
_{\mu }\rangle $ with $\mu =1,2,3,4$. We set the expression of the edge
states as $|\psi _{\mu }\rangle =\left( \psi _{1A},\psi _{1B},\psi
_{1C},\cdots ,\psi _{NA},\psi _{NB},\psi _{NC}\right) $ with $N$ being the
total number of the unit cells. To analytically obtain the wavefunction of
the edge states, we consider the lattice size at the limitation of infinity
large $N\rightarrow \infty $.

For the edge state $|\psi _{1}\rangle $, the component of $|\psi _{1}\rangle
$ at the sublattice $B$ is $\psi _{nB}=0$ with $n$ being the index of the
unit cell in the $y$ direction. The components of $|\psi _{1}\rangle $ in
the first unit cell at the bottom are $\left( \psi _{1A},\psi _{1B},\psi
_{1C}\right) =\left( 1,0,e^{i\phi }\right) $ with $\phi =$sgn$(m)\arccos
(\gamma /m)$. The components of $|\psi _{1}\rangle $ in the $n$-th unit cell
satisfy the recursion relation $\left( \psi _{nA},\psi _{nB},\psi
_{nC}\right) =\rho ^{n-1}\left( \psi _{1A},\psi _{1B},\psi _{1C}\right) $
with $\rho =-e^{-i\phi }-e^{-i(\phi +k_{x})}-1$. The probabilities of edge
states for the sublattices $A$ and $C$ in every unit cell are identical and
decay exponentially from bottom to top of the Lieb lattice as shown in Fig. %
\ref{TNT1}(c). For the edges state $|\psi _{3}\rangle $, we have similar
wavefunction distribution. The components of $|\psi _{3}\rangle $ in the
first unit cell at the bottom are $\left( \psi _{1A},\psi _{1B},\psi
_{1C}\right) =\left( 1,0,e^{-i\phi }\right) $ and decay as the index of the
unit cells at the rate $\rho =-e^{i\phi }-e^{i(\phi -k_{x})}-1$.

\begin{figure}[tb]
\setlength{\abovecaptionskip}{0cm} \centering
\begin{minipage}[!htbp]{0.42\linewidth}
		\setlength{\abovecaptionskip}{0cm} \negthinspace
		\includegraphics[bb=35 305 300 530, width=5cm, angle=0,clip]{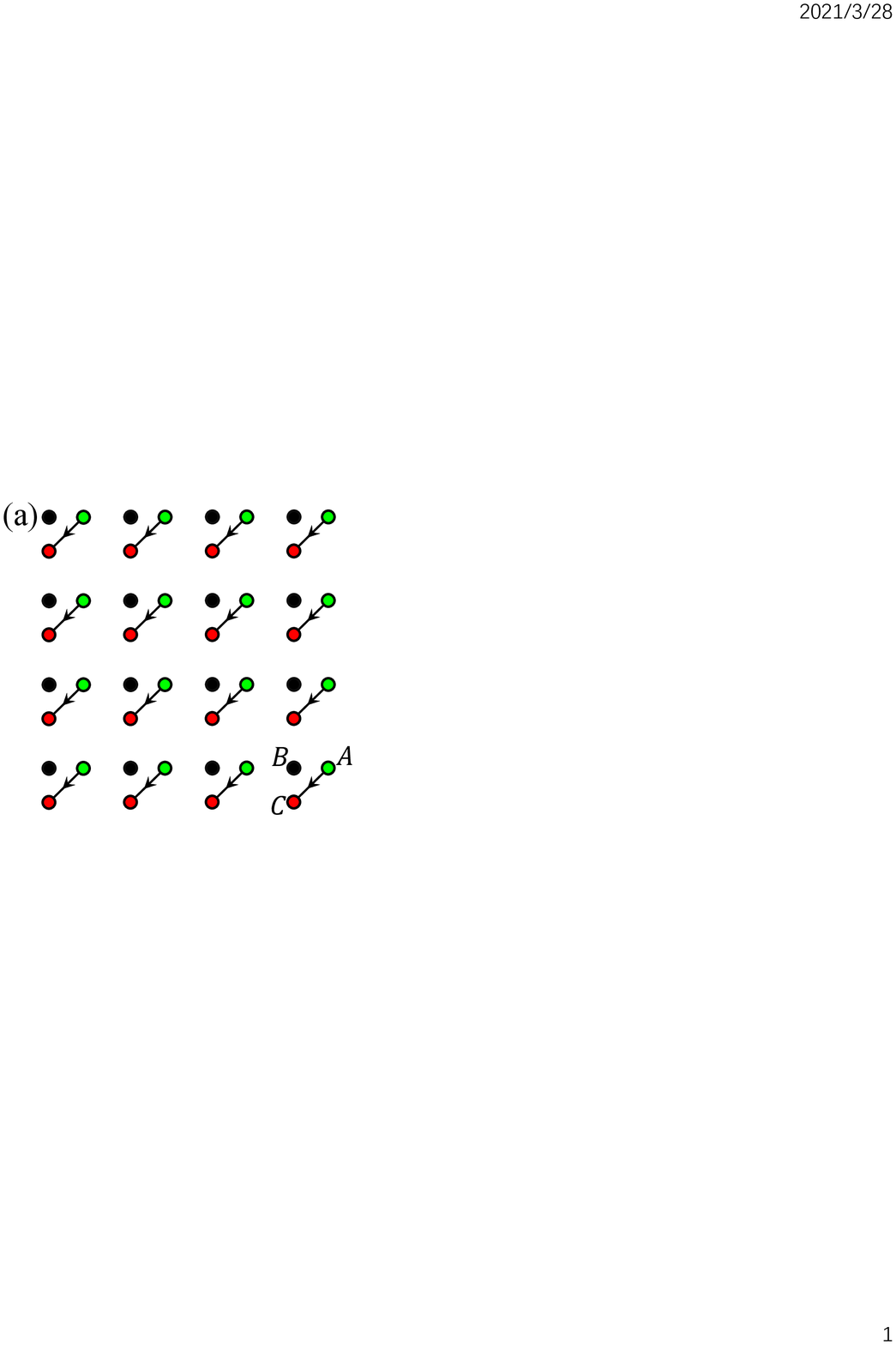}
	\end{minipage}
\hspace{0.5cm}
\begin{minipage}[!htbp]{0.5\linewidth}
		\includegraphics[bb=1 -70 880 800,width=4.5cm,clip]{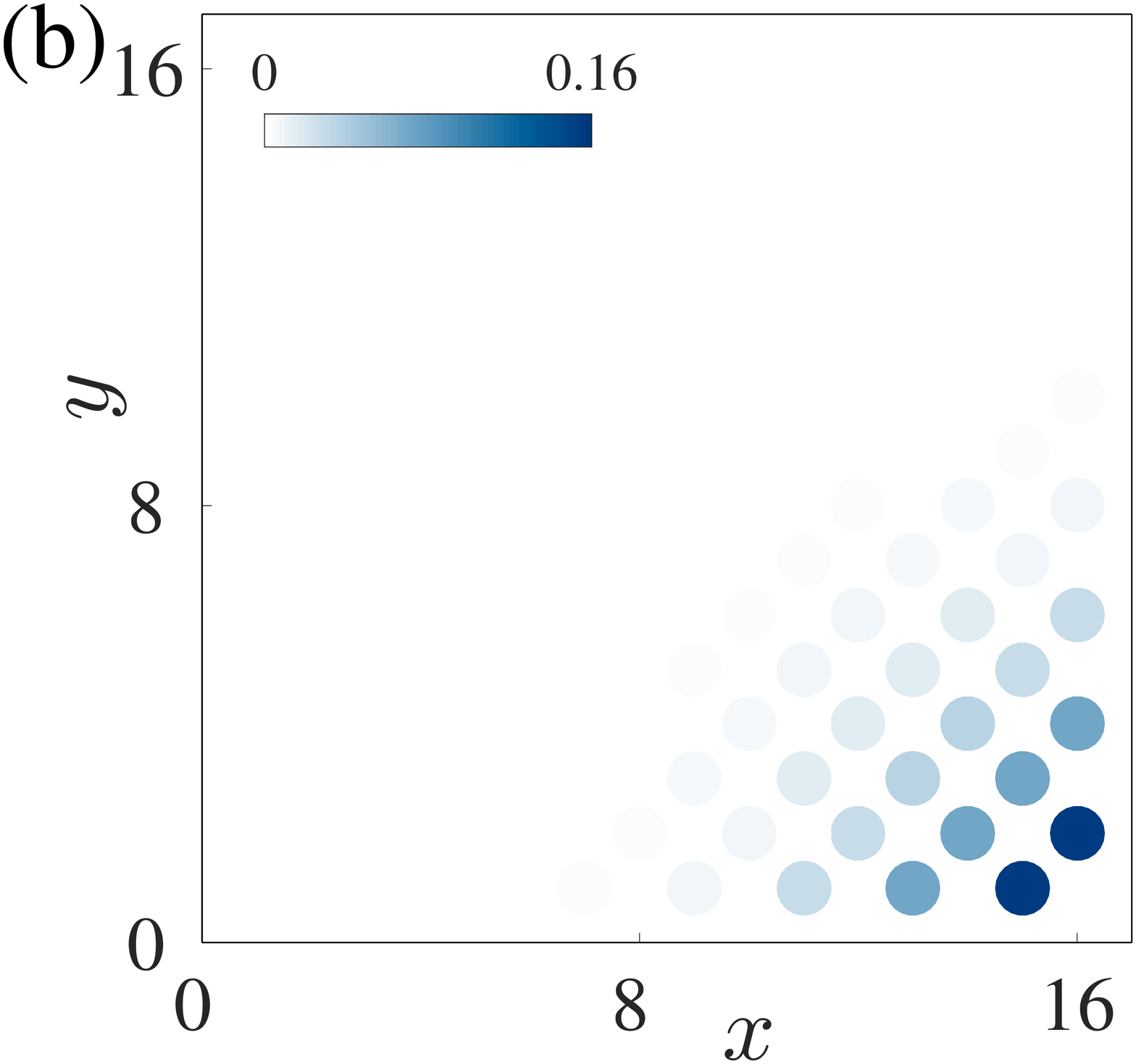}
	\end{minipage}\newline
\caption{Corner states of the Lieb lattice at the energies $E_{1}$ and
$E_{3}$. (a) The dimerized structure of the corner state, being
parity-time-symmetric. (b) The probability distribution of the corner state.
The parameters are $\kappa=1$, $m=-1$, $\protect\gamma=0.1$ and the lattice
consists of 8$\times $8 unit cells.} \label{Corn}
\end{figure}

For the edge state $|\psi _{2}\rangle $, the component of $|\psi _{2}\rangle
$ at the sublattice $C$ vanishes, $\psi _{nC}=0$. The components of $|\psi
_{2}\rangle $ in the first unit cell at the top are $\left( \psi _{NA},\psi
_{NB},\psi _{NC}\right) =\left( e^{i(\phi +k_{x}/2)},1,0\right) $ with $\phi
=\arcsin (\gamma /[4\cos (k_{x}/2)])$ for $m>0$, and $\pi -\arcsin (\gamma
/[4\cos (k_{x}/2)])$ for $m<0$. The components of $|\psi _{2}\rangle $ in
the $n$-th unit cell satisfy the recursion relation $\left( \psi _{nA},\psi
_{nB},\psi _{nC}\right) =\rho ^{N-n}\left( \psi _{NA},\psi _{NB},\psi
_{NC}\right) $ with $\rho =-1-ime^{i(\phi +k_{x}/2)}$. The probabilities of
the edge states for the sublattices $A$ and $B$ in every unit cell are
identical and decay exponentially from top to bottom of the Lieb lattice as
shown in Fig. \ref{TNT1}(d). For the edge state $|\psi _{4}\rangle $, the
result is similar. The components of $|\psi _{4}\rangle $ in the first unit
cell at the top are $\left( \psi _{NA},\psi _{NB},\psi _{NC}\right) =\left(
-e^{i(k_{x}/2-\phi )},1,0\right) $ and decay as the index of the unit cells
at the rate $\rho =-1+ime^{i(k_{x}/2-\phi )}$ with $\phi =\arcsin (\gamma
/[4\cos (k_{x}/2)])$ for $m<0$, and $\pi -\arcsin (\gamma /[4\cos
(k_{x}/2)]) $ for $m>0$.
\begin{figure*}[tbp]
\setlength{\abovecaptionskip}{0.5cm}
\includegraphics[bb=6 -10 1850
	750,width=18cm,height=7.5cm, clip]{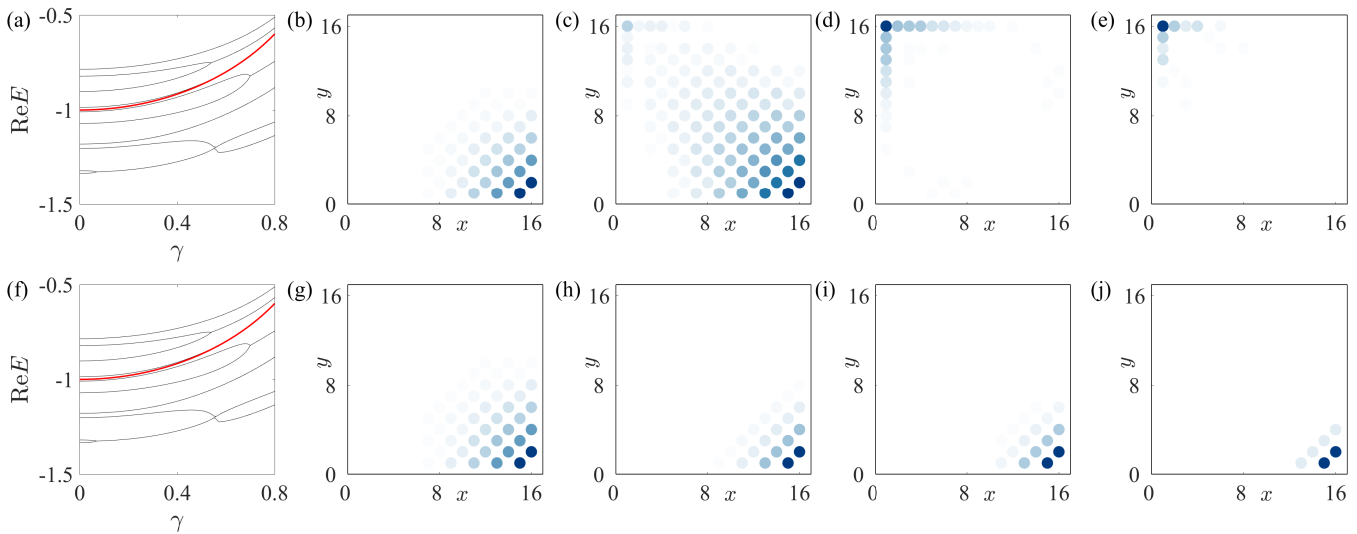}
\caption{The profiles of corner state in the real space for finite-size
non-Hermitian Lieb lattice affected by the non-Hermiticity. The 2D lattice
consists of 8$\times $8 unit cells. (a) and (f) The plots of energy levels
close to the corner states (red line) as functions of $\protect\gamma $. $%
m=1 $ for the upper panels and $m=-1$ for the lower panels. (b, g) $\protect%
\gamma =0$, (c, h) $\protect\gamma =0.4$, (d, i) $\protect\gamma =0.5$, (e,
j) $\protect\gamma =0.8$. The other parameter is $\protect\kappa =1$.}
\label{DCorn}
\end{figure*}

The energy spectra for the lattice under PBC in the $y$ direction and OBC in
the $x$ direction are shown in Figs.~\ref{TNT2}(a)-\ref{TNT2}(b). There are
still four edge states in pairs within the band gap, in the form of%
\begin{eqnarray}
E_{1} &=&\sqrt{m^{2}-\gamma ^{2}}, \\
E_{2} &=&\sqrt{4\kappa ^{2}\cos ^{2}(k_{y}/2)-\gamma ^{2}/4}-i\gamma /2, \\
E_{3} &=&-\sqrt{m^{2}-\gamma ^{2}}, \\
E_{4} &=&-\sqrt{4\kappa ^{2}\cos ^{2}(k_{y}/2)-\gamma ^{2}/4}-i\gamma /2.
\end{eqnarray}%
$E_{1}$ and $E_{3}$ are independent of the momentum $k_{y}$ with $%
E_{3}=-E_{1}$. $E_{2}$ and $E_{4}$ are complex with constant loss rate $%
-i\gamma /2$ with $E_{4}=-E_{2}^{\ast }$. We analyze the wave functions of
the four edge states in the same way as above.

For the edge state $|\psi _{1}\rangle $, the component of $|\psi _{1}\rangle
$ at the sublattice $B$ is $\psi _{\upsilon B}=0$ with $\upsilon $ being the
index of the unit cell in the $x$ direction. The components of $|\psi
_{1}\rangle $ in the first unit cell at the right are $\left( \psi
_{NA},\psi _{NB},\psi _{NC}\right) =(e^{-i\phi },0,1)$ with $\phi =$sgn$%
(m)\arccos (\gamma /m)$. The components of $|\psi _{1}\rangle $ in the $%
\upsilon $-th unit cell satisfy the recursion relation $\left( \psi
_{\upsilon A},\psi _{\upsilon B},\psi _{\upsilon C}\right) =\rho
^{N-\upsilon }(\psi _{NA},\psi _{NB},\psi _{NC})$ with $\rho =-1-e^{i\phi
}-e^{i(k_{y}+\phi )}$. The probabilities of edge states for the sublattices $%
A$ and $C$ in every unit cell are identical and decay exponentially from
right to left of the Lieb lattice as shown in Fig. \ref{TNT2}(c). For the
edges state $|\psi _{3}\rangle $, we have similar wavefunction distribution.
The components of $|\psi _{3}\rangle $ in the first unit cell at the right
are $\left( \psi _{1A},\psi _{1B},\psi _{1C}\right) =(e^{i\phi },0,1)$ and
decay as the index of the unit cells at the rate $\rho =-1-e^{-i\phi
}-e^{i(k_{y}-\phi )}$.

For the edge state $|\psi _{2}\rangle $, the component of $|\psi _{2}\rangle
$ at the sublattice $A$ vanishes, $\psi _{\upsilon A}=0$. The components of $%
|\psi _{2}\rangle $ in the first unit cell at the left are $\left( \psi
_{1A},\psi _{1B},\psi _{1C}\right) =(0,1,e^{-i(\phi +k_{y}/2)})$ with $\phi
=\arcsin (\gamma /[4\cos (k_{y}/2)])$ for $m>0$, and $\pi -\arcsin (\gamma
/[4\cos (k_{y}/2)])$ for $m<0$. The components of $|\psi _{2}\rangle $ in
the $\upsilon $-th unit cell satisfy the recursion relation $\left( \psi
_{\upsilon A},\psi _{\upsilon B},\psi _{\upsilon C}\right) =\rho ^{\upsilon
-1}\left( \psi _{1A},\psi _{1B},\psi _{1C}\right) $ with $\rho
=-1+ime^{-i(\phi +k_{y}/2)}$. The probabilities of the edge states for the
sublattices $B$ and $C$ in every unit cell are identical and decay
exponentially from left to right of the Lieb lattice as shown in Fig. \ref%
{TNT2}(d). For the edge state $|\psi _{4}\rangle $, the result is similar.
The components of $|\psi _{4}\rangle $ in the first unit cell at the right
are $\left( \psi _{NA},\psi _{NB},\psi _{NC}\right) =(0,1,-e^{i(\phi
-k_{y}/2)})$ and decay as the index of the unit cells at the rate $\rho
=-1-ime^{i(\phi -k_{y}/2)}$ with $\phi =\arcsin (\gamma /[4\cos (k_{y}/2)])$
for $m<0$, and $\pi -\arcsin (\gamma /[4\cos (k_{y}/2)])$ for $m>0$.
\begin{figure*}[tbp]
\setlength{\abovecaptionskip}{0.5cm} \includegraphics[bb=6 -10 1000
450,width=18cm,height=8.2cm, clip]{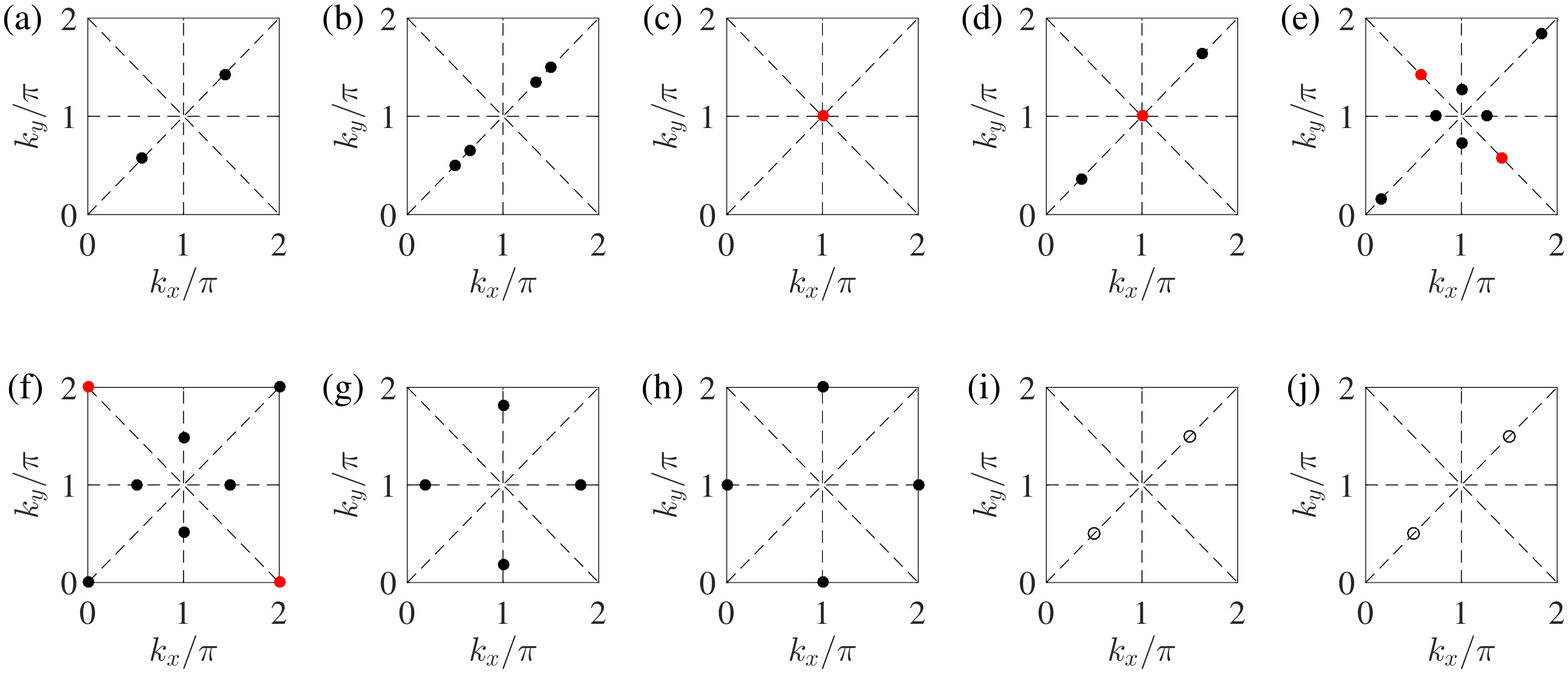}
\caption{Complex energy band structures of the non-Hermitian Bloch
Hamiltonian $h(\mathbf{k})$. The system parameters are determined by the
representative points and traces the variation of EP in different region of
Fig. \protect\ref{PhaseD1}. The EPs projected on the $k_{x}$-$k_{y}$ plane:
(a) $m=1,\protect\gamma =0.388$, (b) $m=1,\protect\gamma =0.5$, (c) $m=0.5,
\protect\gamma =0.5$, (d) $m=1,\protect\gamma =1$, (e) $m=1,\protect\gamma %
=2 $, (f) $m=1,\protect\gamma =3$, (g) $m=1,\protect\gamma =4$, (h) $m=1,
\protect\gamma =4.162$, (i) $m=-\protect\sqrt{2},\protect\gamma =0$ and (j) $%
m= \protect\sqrt{2},\protect\gamma =0$. The other system parameter is $%
\protect\kappa =1$. The solid black dots, red black dots and black circles
represent EP2, EP3 and DP2 respectively. Notably, the variation of $\protect%
\gamma $ can induce the splitting and the merging of the different types of
EP so that the system can exhibit rich structure.}
\label{EP}
\end{figure*}

From Figs. \ref{TNT1}(c) and \ref{TNT2}(c), the edge state energies are
independent of the momentum and the edge states locate on the sublattices $A$
and $C$ for the non-Hermitian Lieb lattice under open boundary in either the
$x$ or the $y$ direction. If the boundaries of the non-Hermitian
Lieb lattice on both the $x$ and the $y$ directions are open, the corner
state appears at the bottom-right of the non-Hermitian Lieb lattice. The
structure of the corner state is schematically illustrated in Fig.~\ref{Corn}%
(a) and the probability distribution of the corner state is shown in Fig.~%
\ref{Corn}(b). The destructive interference plays the crucial role to form the dimerized pattern of the corner state. In the Hermitian case with zero $\gamma $, numerical simulation for the finite system shows that there exist two corner states
with energy $E_{c}=\pm \left\vert m\right\vert $. The corresponding corner states appear at the bottom-right of the lattice as shown in Figs.~\ref%
{DCorn}(b) and \ref{DCorn}(g), where the sublattice $B$ is unoccupied. For
the case with nonzero $\gamma $, the non-Hermitian term affects the corner
states in a subtle way. Numerical results are plotted in Fig.~\ref{DCorn},
the influence of the gain and loss is demonstrated.

We find that (i) for $\gamma /m<0$, the magnetic flux enclosed in the unit
cell is $\pi /2$. The corner states always appear at the bottom-right of the
lattice, and become more localized with the increase of non-Hermiticity $%
\left\vert \gamma \right\vert $. while (ii) for $\gamma /m>0$, the magnetic
flux enclosed in the unit cell is $-\pi /2$. The corner state becomes more
extended as $\left\vert \gamma \right\vert $\ increases, and the
bottom-right corner states slowly evolve into the top-left corner state.\ A
new corner state appears at the top-left of the lattice as $\left\vert
\gamma \right\vert $\ across about $0.5$. Interestingly, the occupation of
the $B$ sublattice is dominant in this case.

\section{Band structure of gapless phase characterized by the EPs}

\label{IV}

In this section, we investigate the band structure of the non-Hermitian Lieb
lattice, where the existence of the EP in the spectrum is featured for the
gapless phase. The EP\ is unique for the non-Hermitian physics and is
associated with the level coalescence, where not only the eigen energies but
also the eigenstates become the same. Many interest effects without
Hermitian counterparts arise around the EP, ranging from the square root
frequency dependence \cite{Peng2014}, the nontrivial topological property
resulting from the Riemann sheet structures \cite{Rotter2015,Zhou2018a}, to
unidirectional reflectionless and coherent perfect absorption \cite%
{Zhang2013,Jin2016,Parto2021}.

To calculate the EP of the non-Hermitian Lieb lattice, we define $\Delta $
as the discriminant\ of the cubic equation (\ref{cubic}). The real and
imaginary parts of $\Delta $ read%
\begin{eqnarray}
\mathrm{Im}(\Delta ) &=&m\gamma h_{x}h_{y}\sin
[(k_{x}+k_{y})/2](h_{y}^{2}-h_{x}^{2}),  \label{d1} \\
\mathrm{Re}(\Delta ) &=&p^{3}/27+\mathrm{Re}(q^{2})/4.  \label{d2}
\end{eqnarray}%
$\Delta =0$ with nonzero $\gamma $ signifies the EP in the spectrum, where
the energy bands are contacted. The phase diagram for the band structure
determined from Eqs. (\ref{d1})-(\ref{d2}) is depicted in Fig. \ref{PhaseD1}%
. The non-Hermitian system is gapless when the energy bands are inseparable.
In the gapless phase, at least two of the bands touch and $\mathbf{k}_{
\mathrm{EP}}$\ is the EP in the BZ; besides, the EPs might present in
different energy bands at the same momentum $\mathbf{k}_{\mathrm{EP}}$. The
gapless phase is divided into three regions in terms of the
types of EPs rather than the exceptional ring \cite{Yoshida2019}, denoted as
yellow, green, and cyan. In the yellow regions, there are four EP2s
(two-state coalescence); six EP2s and two EP3s (three-state coalescence)
present in the green regions; and the cyan regions have four EP2s. The EP
merge or split when the system parameters cross the solid lines as the
boundaries between colored regions.

\begin{figure*}[tbp]
\setlength{\abovecaptionskip}{0.5cm}  \includegraphics[bb=25 0 1600
540,width=18.8cm,height=7cm, clip]{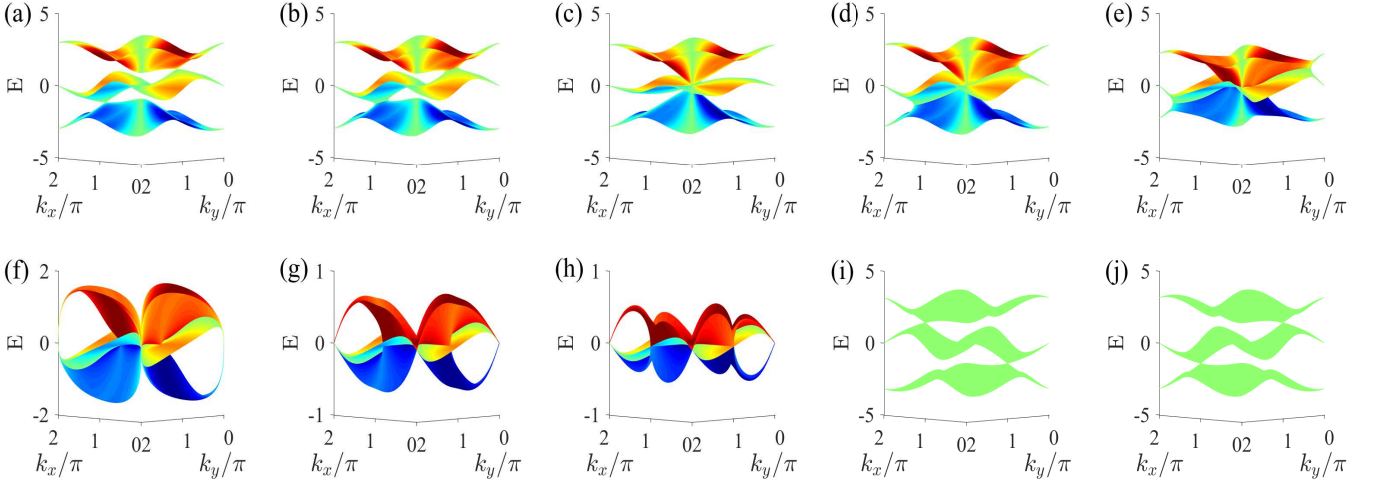}
\caption{Energy bands structures of the non-Hermitian Hamiltonian in the
momentum space as the counterpart of Fig. \protect\ref{EP}. In all panels,
the vertical axis represents the real part of eigenvalues and the color
indicates the imaginary part of eigenvalues. (a) $m=1,\protect\gamma =0.388$%
, (b) $m=1,\protect\gamma =0.5$, (c) $m=0.5, \protect\gamma =0.5$, (d) $m=1,%
\protect\gamma =1$, (e) $m=1,\protect\gamma =2 $, (f) $m=1,\protect\gamma =3$%
, (g) $m=1,\protect\gamma =4$, (h) $m=1, \protect\gamma =4.162$, (i) $m=-%
\protect\sqrt{2},\protect\gamma =0$ and (j) $m= \protect\sqrt{2},\protect%
\gamma =0$. The other system parameter is $\protect\kappa =1$.}
\label{EB}
\end{figure*}

We take $\kappa =1$ as an illustration and analytically determine the EPs
and the boundaries within the gapless phases from Im$(\Delta )=0$ and Re$%
(\Delta )=0$. The EPs may appear at $k_{x}=\pi $, $k_{y}=\pi $, and $%
\left\vert k_{x}\right\vert =\left\vert k_{y}\right\vert $.

(i) The EP2s appear at $k_{x}=k_{y}$, we obtain $h_{x}^{2}=h_{y}^{2}$ and $%
\mathrm{Im}(\Delta )=0$. The cyan and green regions in the phase diagram
have EP2s in this case, where the energies at the EP2s are real with $%
\mathrm{Im}(E)=0$. From Re$(\Delta )=0$, we obtain
\begin{equation}
4(\gamma ^{2}-m^{2}-2h_{x}^{2})^{3}=27h_{x}^{6}m^{2}(h_{x}^{2}-4),
\label{B4}
\end{equation}%
which determines the green and cyan regions in Fig. \ref{PhaseD1}. The
purple boundary and the boundary $\left\vert \gamma \right\vert =\left\vert
m\right\vert $ in red enclose the cyan region with four EP2s, but the blue
boundary and the boundary $\left\vert \gamma \right\vert =\left\vert
m\right\vert $ enclose the green region with two EP2s. The representative
points in the two regions\ are shown in Fig. \ref{EP}(b) and Fig. \ref{EP}%
(e) respectively. We notice that on the dashed line $k_{x}=k_{y}$, the
former has four EP2s, and the latter has only two. At the boundary, the
number of EP2s will be reduced due to merging. In Fig. \ref{EP}(a) and\ Fig. %
\ref{EP}(d), we notice that the four EP2s in the cyan region merge into two
at the purple or red boundaries. In Fig. \ref{EP}(c) and\ Fig. \ref{EP}(f),
we notice that the two EP2s in the green region merge into one EP at the
blue or black boundaries.

(ii) The EP3s appear at $k_{x}=-k_{y}$, we obtain $\sin (k_{x}+k_{y})/2=0$
and $\mathrm{Im}(\Delta )=0$. From Re$(\Delta )=0$, we obtain%
\begin{equation}
\gamma ^{2}-m^{2}=2h_{x}^{2},  \label{B2}
\end{equation}%
which determines the green region. The boundary $\gamma ^{2}-m^{2}=8$ in
blue and the boundary $\left\vert \gamma \right\vert =\left\vert
m\right\vert $ enclose the green region with two EP3s. This indicates that
the green region has EP3s in addition to EP2s. The EP3s are the red dots in
Fig.~\ref{EP}(e), where the three band coalesce at zero energy $E=0$. Two
EP3s merge to one at $(k_{x},k_{y})=(0,0)$ in Fig.~\ref{EP}(f) at the blue
boundary and one at $(k_{x},k_{y})=(\pi ,\pi )$ in Figs.~\ref{EP}(c)-(d) at $%
\left\vert \gamma \right\vert =\left\vert m\right\vert $. The non-Hermitian 
Lieb lattice does not hold the anti-unitary symmetries and the EP3s are not topological stable \cite{Delplace2021,Kkawab2019,TBes2019}.

(iii) The EP2s appear at $k_{x}=\pi $ or $k_{y}=\pi $, which leads to $%
h_{x}=0$ or $h_{y}=0$ and $\mathrm{Im}(\Delta )=0$. From Re$(\Delta )=0$, we
obtain
\begin{equation}
4(\gamma ^{2}-m^{2}-h_{x}^{2})^{3}=27\gamma ^{2}h_{x}^{4},  \label{B1}
\end{equation}%
which determines the green and yellow regions of the phase diagram in Fig. %
\ref{PhaseD1}. The boundary $(\gamma ^{2}-m^{2}-4)^{3}=108\gamma ^{2}$ in
green and the boundary $\left\vert \gamma \right\vert =\left\vert
m\right\vert $ enclose these two regions with four EP2s. At these four EP2s,
the coalesced band energies are imaginary with Re$(E)=0$. The representative
configurations are shown in Figs.~\ref{EP}(d)-\ref{EP}(h), showing the
movement and merging of EPs. Four EP2s move along the two dashed lines $%
k_{x}=\pi $ and $k_{y}=\pi $ in Figs.~\ref{EP}(e)-\ref{EP}(g) until they
merge into two EP2s at $(k_{x},k_{y})=(0,\pi ),(\pi ,0)$ at the green
boundary in Fig. \ref{EP}(h) or merge as one EP at $(k_{x},k_{y})=(\pi ,\pi )
$ at $\left\vert \gamma \right\vert =\left\vert m\right\vert $ in Fig. \ref%
{EP}(d). The energy bands for Fig.~\ref%
{EP} are shown in Fig.~\ref{EB}, where the bulk Fermi-arc is observed.

In the phase diagram, the cyan region has four EP2s at $k_{x}=k_{y}$; the
green region has two EP2s at $k_{x}=k_{y}$, two EP3s at $k_{x}=-k_{y}$, and
four EP2s at $k_{x}=\pi $ and $k_{y}=\pi $; the yellow region has four EP2s
at $k_{x}=\pi $ and $k_{y}=\pi $. Under the OBC, the edge states exist 
in the gapless region in cyan; they are the remnant of topological features inherited from the Hermitian Lieb lattice; however, the edge state does not exist in the gapless regions in green and yellow. At even larger gain and loss, the non-Hermitian Lieb lattice enters the trivial phase.

\section{Conclusion}

\label{V}To conclude, we have proposed an anisotropic 2D non-Hermitian Lieb
lattice, which has gain and loss along the horizontal and vertical
directions, respectively. The nonreciprocal intra-cell coupling creates
non-trivial topology. The gain and loss result in active and dissipative
topological edge states with net gain and net loss, respectively. The
non-Hermitian Lieb lattice also supports passive topological edge states.
Interestingly, the gain and loss can alter the localization position of the
corner state. When the magnetic flux enclosed in the unit cell is $\pi /2$,
the corner states are always in one corner of the lattice; however, when the
magnetic flux enclosed in the unit cell is $-\pi /2$, the corner states
slowly evolve into the opposite corner as the non-Hermiticity increases. The
interplay between the magnetic flux and non-Hermiticity produces rich band
structures featured from different types of EPs. The topological properties
and rich band structures of the non-Hermitian Lieb lattice benefit our
understanding of three-band lattice and robust non-Hermitian transport.

\section*{Acknowledgement}

We acknowledge the support of the National Natural Science Foundation of
China (Grants No. 11975128, No. 11975166, and No. 11874225). X.Z.Z. is also
supported by the Program for Innovative Research in University of Tianjin
(Grant No. TD13-5077).


\end{document}